\documentclass[seceq,preprint]{ptptex}

\newcommand{\tr}{\mathop{\rm tr}\nolimits}
\newcommand{\Tr}{\mathop{\rm Tr}\nolimits}
\newcommand{\rmd}{{\rm d}}
\newcommand{\n}{\nonumber}
\newcommand{\U}{\mathop{\rm {}U}}
\newcommand{\sgn}{\mathop{\rm sgn}}
\newcommand{\vol}{\mathop{\rm vol}}

\preprintnumber[3cm]{KEK-TH-861}

\title{
Chiral Anomaly 
and Ginsparg-Wilson Relation on the 
Noncommutative Torus%
}

\author{%
Satoshi \textsc{Iso}\footnote{E-mail: satoshi.iso@kek.jp}
and Keiichi \textsc{Nagao}\footnote{E-mail: nagao@post.kek.jp} 
}

\inst{
Institute of Particle and Nuclear Studies, \\
High Energy Accelerator Research Organization (KEK), 
Tsukuba 305-0801, Japan 
}

\abst{
We evaluate chiral anomaly on the noncommutative torus 
with the overlap Dirac operator satisfying the 
Ginsparg-Wilson relation in arbitrary even dimensions. 
Utilizing a topological argument we show that the 
chiral anomaly is combined into a form of the Chern character 
with star products.
}

\begin{document}

\maketitle

\section{Introduction}
Various matrix models have been proposed as a nonperturbative
formulation of the superstring theory. Type IIB matrix model
is one of such proposals \cite{IKKT} 
and various properties have been 
investigated. In particular, a possibility of the
dynamical generation of four dimensional space-time 
has been examined from various view points 
\cite{spacetime}.
A connection to the noncommutative field theory \cite{Connes}
has also been studied \cite{CDS,Li,AIIKKT,Ambjorn} and several
nontrivial dynamics were discussed in connection 
with string theory \cite{NCYM}.
There are still many unsolved issues such as the 
embedding of curved space-time \cite{metric},
locality of the field theories
on the dynamically generated space-time or
the origin of four dimensional chiral fermions.
Also, we need to understand how to describe
global topology of space-time and gauge
configurations within finite dimensional
matrix models.
\par
A possible origin of four dimensional chiral fermions 
using orbifold matrix models was
discussed in \cite{AIS} but this approach is 
not yet completely satisfactory since we need to restrict 
the degrees of
freedom by hand. Another approach to define chiral fermions in
four dimensions will be to mimic the Kaluza-Klein compactification
with non-trivial indices.
In the paper \cite{AIN2}, we have proposed to use the Ginsparg-Wilson (GW)
relation \cite{GW}
to define chiral structures in finite dimensional
matrix models with general curved backgrounds and 
showed that this construction makes it possible
to define a topological invariant for  gauge field configurations
with only finite number of degrees of freedom.
Furthermore we gave an example on the fuzzy two-sphere 
in general background gauge fields
and constructed  chirality and
Dirac operators 
satisfying the GW relation. We showed that the 
topological invariant coincides with the first Chern
class in the commutative limit. 
\par
We can apply the same technique to classify the topological 
structures of the gauge field 
configurations on noncommutative tori.
The GW relation was first introduced in finite matrix models with 
the background of a noncommutative torus in ref. \cite{NV}. 
In this paper, based on the formulation of ref. \cite{NV}, 
we evaluate the topological charge density for the 
overlap Dirac operator on the noncommutative torus and 
show that it becomes the star generalization of the Chern class.
\par
In lattice gauge theory the formalism based on the GW relation 
has been investigated recently. The first important observation was 
that, in the presence of the mass defect which is introduced 
as a scalar background in higher $4+x$ dimensions, a chiral 
fermion appears at the defect. The topological defects are 
a kink for $x=1$ and a vortex for $x=2$.
So far a domain wall fermion ($x=1$)~\cite{dwfermion} 
and a vortex fermion ($x=2$)~\cite{Nagao:2001tc} 
are constructed on the lattice. {}From the former model 
a practical solution to the GW relation is 
obtained~\cite{Neuberger:1998fp}.
This is the overlap Dirac operator.
Anomaly free abelian chiral gauge theory is also constructed 
on the lattice by using the GW relation~\cite{Luscher:1998du}.
In the ordinary lattice gauge theory,
chiral anomaly for the overlap Dirac operator 
has been studied in four 
dimensions~\cite{Kikukawa:1998pd} 
and in arbitrary even dimensions~\cite{Fujiwara:2002xh}.
The form of the Chern character is derived by  
cohomological arguments at a finite lattice 
spacing~\cite{Luscher:1999kn}. 
\par
In this paper, we evaluate the chiral anomaly 
on arbitrary even dimensional noncommutative torus 
for the noncommutative version \cite{NV}
of the overlap 
Dirac operator~\cite{Neuberger:1998fp} 
in the continuum limit. 
The chiral anomaly is evaluated as an integral of the 
Chern character and the only difference from the ordinary
anomaly is that the product of gauge fields is replaced by
the star product on noncommutative torus. 
Anomaly on $2$-dimensional noncommutative torus is 
also calculated in ref.~\cite{Correa:2001rd}
\par
A particular importance to consider 
the GW relation and  topological
invariants on the noncommutative torus will be that
we can 
compare the various properties with those in the ordinary
lattice gauge theories that have been intensively
studied.
They include the locality condition, 
the admissibility condition \cite{Hernandez:1998et} and
the classification of the admissible gauge field 
configurations \cite{Luscher:1998du}. 
We want to discuss them in a separate paper. 
Parity anomaly on noncommutative torus is also 
evaluated in ref.~\cite{Nishimura:2002hw}.
\par
In section 2, we give a brief review of the noncommutative torus
and in section 3, the overlap Dirac operator is introduced
on the noncommutative torus. Section 4 is the main part 
of the paper and
the chiral anomaly, that is, the topological charge density
is evaluated. In the appendix we give a detailed evaluation 
of the coefficient of the anomaly 
following ref.\cite{Fujiwara:2002xh}.

\section{Noncommutative Torus}
A noncommutative torus is one of the simplest examples
of noncommutative geometries which can be realized
in terms of finite matrices. 
More precisely, a complete basis of wave functions 
on noncommutative torus forms a finite dimensional
matrix algebra\cite{CDS,Ambjorn}.
\par
We use the following specific matrix 
representation of the $d$-dimensional ($d$ is an even integer) 
noncommutative torus.
First we introduce the 't Hooft matrices
$U$ and $V$. They are 
$L\times L$-dimensional matrices defined by
\begin{equation}
U=
\begin{pmatrix}
 0&1& & &0\cr 
& 0 & 1 & & \cr 
& &\ddots & \ddots & \cr & & & \ddots & 1 
\cr 1& & & &0\cr
\end{pmatrix}
 ~~~~~,~~~~~
V=\begin{pmatrix} 
1& & & & \cr & \omega & & & \cr 
& & \omega^2 & & \cr 
& & &\ddots& \cr 
& & & & \omega^{L-1} \cr 
\end{pmatrix}
,
\label{clockshift}
\end{equation}
where $\omega=e^{i\frac{2\pi}{L}}$. 
They are unitary and satisfy the following identities,
\begin{equation}
U^L=1, \quad V^L=1, \quad UV=e^{i\frac{2\pi}{L}}VU .
\end{equation}
$\exp(-i \pi nm/L) U^n V^m $ forms a complete basis
of $L \times L$ matrices and 
any hermitian matrices can be expressed as a sum of them.
\par
A $d$-dimensional 
noncommutative torus can be constructed by 
taking a $d/2$ tensor product of these matrices.
First we define $N \times N$ matrices by 
\begin{eqnarray}
Z_{2i-1}&=&{\bold 1}_{L}\otimes\cdots\otimes U
\otimes\cdots\otimes{\bold 1}_{L}, \label{Z1} \\
Z_{2i} &=& {\bold 1}_L\otimes\cdots\otimes 
V \otimes\cdots\otimes{\bold 1}_L \label{Z2} 
\end{eqnarray}
where $i=1, \cdots, d/2$ and $N=L^{d/2}$. 
On the right hand side of eqs.(\ref{Z1})(\ref{Z2}), 
$U$ and $V$ are in the i-th slot respectively. 
They satisfy the following relation,
\begin{equation}
Z_\mu Z_\nu = e^{-2\pi i \Theta_{\mu\nu}} Z_\nu Z_\mu,
\end{equation}
where $\Theta_{\mu\nu}=\epsilon_{\mu \nu}/L$.
Here 
$\epsilon_{\mu \nu}$ is a $d \times d$ 
skew diagonal matrix which has the following form,
\begin{equation}
\epsilon_{\mu \nu}=
\begin{pmatrix}
0 & -1 & & & \cr 
1 & 0 & & & \cr 
& & \ddots & & \cr 
& & & 0 & -1 \cr 
& & & 1 & 0 \cr
\end{pmatrix}
.
\end{equation}
These $Z_{\mu}$ 
are building blocks of wave functions on the
noncommutative torus and any $N \times N$ hermitian
matrix can be expressed in terms of these matrices.
For notational simplicity, we formally introduce
hermitian noncommutative coordinates $\hat{x}_{\mu}$
as $Z_{\mu}=\exp(2 \pi i \hat{x}_{\mu}/L_{phys})$.
They satisfy
\begin{equation}
\left[\hat x_\mu,\hat x_\nu \right]=i\theta_{\mu\nu}
\end{equation}
where
$\theta_{\mu\nu}=\frac{{L_{phys}}^2}{2\pi} 
\Theta_{\mu\nu}$.
We further define a ``lattice spacing'' $a$ by the relation
$L_{phys}=L a$.
\par
Any $N \times N$ hermitian matrix can be expanded as
\begin{equation} 
F= \sum_{\vec m \in ({\bold Z}^d )_L} f_{\vec m} 
(Z_1)^{m_1} (Z_2)^{m_2} \cdots (Z_d)^{m_d}
\exp\left(\pi i \sum_{\mu < \nu} \Theta_{\mu\nu}m_\mu m_\nu\right)
\end{equation}
where 
$\vec m \in ({\bold Z}^d)_L$ means a summation over 
all integral vectors 
$\vec m \in {\bold Z}$ modulo $L$.
With the phase $\exp\left(\pi i \sum_{\mu < \nu} \Theta_{\mu\nu}m_\mu m_\nu\right)$, 
hermiticity of $F$ leads to a condition that 
$f_{\vec m}^*=f_{-\vec m + \vec L}$ where 
$\vec L =(L, L, \cdots, L)$. 
We define noncommutative plane waves $\hat\phi_k$ by 
\begin{equation}
\hat\phi_k=(Z_1)^{m_1} (Z_2)^{m_2} \cdots (Z_d)^{m_d}
\exp\left(\pi i \sum_{\mu < \nu} \Theta_{\mu\nu}m_\mu m_\nu\right)
\end{equation}
where $k_\mu=2\pi m_{\mu}/L_{phys}$ is inside the Brillouin 
zone ${\bold B_L}= \left[ -\pi/a, \pi/a \right]$. 
By using $\hat x_\mu$, $\hat\phi_k$ can be simply written as 
$\hat\phi_k=e^{ik\cdot \hat x}=\exp\left(2\pi i \vec m \cdot \vec
{\hat x}/L_{phys}\right)$ and corresponds to taking Weyl orderings. 
$\hat x_\mu$ always appear in the form of $\hat \phi_k$. 
%
The total number of basis wave functions is 
$L^d=N^2$. 
$\hat\phi_{k}$ forms orthogonal basis on the noncommutative 
torus, 
\begin{equation}  
\hat\phi_{k} = e^{ik\cdot \hat x},
\ \ 
\frac{1}{N} \tr \hat\phi_{k}^\dag \hat\phi_{l} = 
\delta_{{k},{l}}.
\end{equation}

\par
Following the paper~\cite{Ambjorn},
we can introduce a ``delta''-function 
\begin{equation}
\Delta(x)=\sum_{\vec m \in ({\bold Z}^d)_L} \hat\phi_k
e^{-2\pi i m \cdot x/{L_{phys}}} .
\end{equation}
In order to keep the periodicity in the momentum 
space $\{\vec m\}$, 
the value of the coordinates $x_{\mu}$ should be 
quantized as $a$ times an integer. Furthermore,
since $m_\mu$ is also quantized as an integer,
$x_\mu$ and $x_\mu+La=x_\mu+L_{phys}$ can be identified. 
Hence, $\vec{x}$ forms a $d$-dimensional lattice 
$\Lambda_L$ where
\begin{equation}
\Lambda_L=\left\{ (x_1,x_2,\cdots,x_d)~\vert~ x_\mu=
a n_\mu, \  n_\mu \in ({\bold Z^d})_L \right\}.
\end{equation}
These delta-functions satisfy the following identities, 
\begin{eqnarray}
\tr \Delta (x) &=& N, \\
\sum_{x \in \Lambda_L} \Delta (x) &=& N^2 {\bold 1}_N, \\
\frac{1}{N}\tr [\Delta(x)\Delta(y)]&=& N^2 \delta_{x,y}.
\end{eqnarray}
By using them, the hermitian matrix
$F$ has a coordinate representation as
\begin{eqnarray}
F &=& \frac{1}{N^2} \sum_{x\in\Lambda_L} 
{\cal F}(x)\Delta(x), \\
{\cal F}(x)&=& \frac{1}{N}\tr[F \Delta(x)]=
\sum_{\vec m \in ({\bold Z^d})_L} f_{\vec{m}} \exp(2 \pi i 
\vec{m}\cdot\vec{x}/L_{phys}) .
\end{eqnarray}
A product of two matrices $F$ and $G$ is mapped 
to a star product of functions 
\begin{eqnarray}
{\cal F}(x)\star {\cal G}(x) &\equiv& \frac{1}{N}
\tr\left[ F G \Delta (x) \right] \nonumber \\
&=&\frac{1}{N^2}\sum_{y\in \Lambda_L}
\sum_{z\in \Lambda_L} {\cal F}(y) {\cal G}(z) 
e^{-2i (\theta^{-1})_{\mu\nu}(x_\mu -y_\mu)(x_\nu -z_\nu)}
\end{eqnarray}
and a trace over matrices is mapped to a summation 
over the lattice
\begin{equation}
\tr F = \frac{1}{N} \sum_{\vec{x} \in \Lambda_L}{\cal F}(\vec{x}).
\end{equation}
\par
In order to define lattice derivatives, 
we introduce shift operators which satisfy
\begin{equation}
\Gamma_\mu \Delta(x)\Gamma_\mu^\dag = \Delta(x-a\hat\mu)
\end{equation}
and hence
\begin{equation}
{\cal F}(x+a\hat\mu)=\frac{1}{N}
\tr \left[\Gamma_\mu F \Gamma_\mu^\dag \Delta (x)\right].
\end{equation}
Here $\hat\mu$ is a unit vector $(0,\cdots,1,\cdots,0)$ which has 
a non-vanishing element in the $\mu$-th direction. 
These shift operators can be constructed explicitly
as 
\begin{eqnarray}
\Gamma_{2i-1}&=&{\bold 1}_{L}\otimes\cdots\otimes V
\otimes\cdots\otimes{\bold 1}_{L},\label{Gamma1}\\
\Gamma_{2i} &=&{\bold 1}_L\otimes\cdots\otimes 
U^\dag
\otimes\cdots\otimes{\bold 1}_L \label{Gamma2},  
\end{eqnarray}
for $i=1,\cdots,\frac{d}{2}$. On the right hand side 
of eqs.(\ref{Gamma1})(\ref{Gamma2}), 
$V$ and $U^\dag$ are in the i-th slot respectively.

\section{Ginsparg-Wilson Fermions on Noncommutative Torus}
In lattice gauge theory, the overlap Dirac operator 
is a practical solution to the GW relation. 
This Dirac operator does not have species doubling for 
some region of parameters $m_0$ and $r$, 
while it has a modified chiral symmetry at a finite 
lattice spacing. 
\par
Following the notation in ref.\cite{AIN2}, 
we first introduce
a chirality operator $\hat{\gamma}$, in addition 
to the ordinary chirality operator $\gamma_{d+1}$, as 
\begin{equation}
\hat{\gamma}= \frac{H}{\sqrt{H^2}}
\end{equation}
and the GW Dirac operator by
\begin{equation}
D_{GW}={1\over a}
\left[1-\gamma_{d+1} \hat{\gamma} \right].
\end{equation}
This operator was first introduced in ref.\cite{NV} 
to define chiral fermions on the noncommutative lattice. 
Here we can take the hermitian operator $H$ as 
the same form as the ordinary overlap Dirac operator 
\cite{Neuberger:1998fp} on the commutative torus : 
\footnote{Our notations: Greek letters,
$\mu$, $\nu$, \dots\ run from $1$ to~$d=2n$. Repeated indices are understood to
be summed over, unless noted otherwise.
$\{\gamma_\mu,\gamma_\nu\}=2\delta_{\mu\nu}$,
$\gamma_\mu^\dagger=\gamma_\mu$
and~$\gamma_{d+1}=(-i)^n\gamma_1\cdots\gamma_d$;
$\gamma_{d+1}^2=1$ and~$\gamma_{d+1}^\dagger=\gamma_{d+1}$ follow from this.}
\begin{eqnarray}
&& H = \gamma_{d+1}(m_0-aD_{\rm w} ),\label{one}\\
&&D_{\rm w}={1\over2}\left[\gamma_\mu(\nabla_\mu^*+\nabla_\mu)
-ar\nabla_\mu^*\nabla_\mu\right],\label{two}
\end{eqnarray}
where $m_0$ and~$r$ are free parameters. 
In the absence of the gauge field, the Dirac operator is 
free of species doubling if $0<\frac{m_0}{r}<2$. 
$\nabla_\mu$ and $\nabla_\mu^*$ are forward
and backward covariant difference operators respectively. 
They are operators acting on matrices defined by 
\begin{eqnarray}
\nabla_\mu\psi&=&{1\over a}
\left[U_\mu \Gamma_\mu (\Gamma_\mu^\dag)^R -1 \right] \psi,\\
\nabla_\mu^*\psi&=&{1\over a}
\left[1-\Gamma_\mu^\dag U_\mu^\dag \Gamma_\mu^R \right] \psi.
\end{eqnarray}
Here the superscript $R$ means that the operator
acts on matrices from the right.
$U_{\mu}$ are analogues of link variables in
lattice gauge theories and $\psi$ is a fermion
in the fundamental representation of the gauge group. 
All these variables $U_\mu$, $\Gamma_\mu$ and $\psi$ are 
$N\times N$ matrices. By using the mapping rules, 
explained in the previous section, from a hermitian matrix 
$F$ to a field on noncommutative lattice ${\cal F}(x)$, 
$\nabla_\mu$ becomes a covariant derivative on noncommutative 
lattices. But here, we use matrix formulation instead of 
explicitly using noncommutative lattices. 
The gauge group is assumed to be abelian
in the following merely for notational simplicity
but fields on noncommutative
geometry are already noncommutative and 
the following calculations can be straightforwardly 
applied to the nonabelian gauge group.  
Due to the property
\begin{eqnarray}
\Gamma_\mu e^{i k_\rho \hat x_\rho}\Gamma_\mu^\dag &=& 
e^{ik_\rho \hat x_\rho} e^{i k_\mu a}, 
\end{eqnarray}
we can write a product of $\Gamma_\mu$ and 
$(\Gamma_\mu^\dag)^R$ as  
\begin{equation}
\Gamma_\mu (\Gamma_\mu^\dag)^R = e^{i\hat p_\mu a},
\end{equation}
where $\hat p_\mu$ is defined to be 
an operator which picks up 
all the momenta of plane waves sited rightward.
Note that the l.h.s gives the definition of the r.h.s, 
which is an operator acting on matrices. This rewriting 
in terms of $\hat p_\mu$ makes the calculation 
in section $4$ simpler and 
similar to that of the ordinary lattice gauge theory. 
%
\par
Under the gauge transformation, 
$D_{GW}$ transforms covariantly 
since $\psi$ and $U_\mu$ transforms as follows,
\begin{eqnarray}
\psi &\rightarrow& g \psi , \label{gpsi}\\
U_\mu &\rightarrow& g U_\mu \Gamma_\mu g^\dag 
\Gamma_\mu^\dag ,\\
\nabla_\mu \psi &\rightarrow& g \nabla_\mu \psi , \\
\nabla_\mu^* \psi &\rightarrow& g \nabla_\mu^* \psi. 
\label{gnablapsi}
\end{eqnarray}

It is the most important property of 
the overlap Dirac operator that it satisfies the Ginsparg-Wilson
relation 
$\gamma_{d+1}D_{GW}+D_{GW} \hat{\gamma}=0$
~\cite{GW}.
Due to this relation, the fermion 
action~$S_{\rm F}=\Tr\bar\psi D_{GW} \psi $ 
is invariant 
under the modified chiral 
transformation~\cite{Luscher, Nieder,AIN2}
\begin{equation}
\delta\psi=\lambda \hat\gamma \psi,\qquad
\delta\bar\psi =\bar\psi \lambda \gamma_{d+1}.
\label{five}
\end{equation}
We note here that $\lambda$ must
transform covariantly as 
\begin{equation}
\lambda \rightarrow g \lambda g^\dag \label{glambda}
\end{equation}
under the gauge transformation.\footnote{
In noncommutative field theory 
there are two different ways to 
define local chiral transformations due to the 
ordering ambiguity of functions and 
the associated chiral currents have also two
different forms~\cite{anomaly}.
One is the invariant current and the other is the covariant 
current, which transforms invariantly and covariantly 
under the gauge transformation respectively. 
In this paper we only consider the covariant type since 
there is no local expression of Chern characters 
for the invariant type.}
\par
The fermion integration measure, 
however, acquires a non-trivial 
jacobian under the transformation
and this gives the chiral anomaly~\cite{Luscher}
\begin{equation}
\delta \rmd\psi \rmd\bar\psi
=-2 q(\lambda) \rmd\psi \rmd \bar\psi,
\end{equation}
where
\begin{equation}
q(\lambda)=\frac{1}{2}{\bold{Tr}}
\left[ \lambda^L \hat\gamma +\lambda^L \gamma_{d+1} \right].
\end{equation}
Here ${\bold{Tr}}$ means a trace over both operators
acting on matrices and $\gamma$-matrices.
In the lattice gauge theory 
this gives the correct chiral Ward-Takahashi identity 
even for finite lattice spacings, so it is especially suitable 
for a study of phenomena related 
to the axial anomaly, such as the $\U(1)$
problem~\cite{Chandrasekharan:1998wg}.
 
\section{Chiral Anomaly and Topological Charge}
We now evaluate the topological charge density
$q(\lambda)$ in a weak coupling expansion.
Hence we write the link variable $U_{\mu}$
as $U_{\mu} = \exp(iaA_{\mu})$ and expand $q(\lambda)$ 
in terms of the gauge field $A_{\mu}$.
First we expand $H$ and $H^2$ as  
\begin{eqnarray}
H&=&H_0 + H_1 + H_2 +H_3+ \cdots ,  \\
H^2 &=& (H^2)_0 +(H^2)_1 +(H^2)_2 +(H^2)_3 + \cdots ,
\end{eqnarray}
where $H_i$ and $(H^2)_i ~(i=0,1,2,\cdots)$ means 
the i-th order term of the gauge field $A_\mu$.
$(H^2)_i$ can be written as
\begin{eqnarray}
(H^2)_0 &=& H_0^2 , \\
(H^2)_1 &=& \left\{H_0, H_1 \right\} ,\\
(H^2)_2 &=& \left\{H_0, H_2 \right\}+H_1^2 ,\\
(H^2)_3 &=& \left\{H_0, H_3 \right\}+ \left\{H_1, H_2 \right\},\\
&\vdots & . \n
\end{eqnarray}
By noting 
\begin{eqnarray}
&&\nabla_\mu+\nabla_\mu^* =
\frac{1}{a} \left( U_\mu \Gamma_\mu (\Gamma_\mu^\dag)^R
-\Gamma_\mu^\dag U_\mu^\dag \Gamma_\mu^R \right) , \\
&&\nabla_\mu^* \nabla_\mu =
\frac{1}{a}(\nabla_\mu - \nabla_\mu^*)
=\frac{1}{a^2}
\left(U_\mu \Gamma_\mu (\Gamma_\mu^\dag)^R 
+ \Gamma_\mu^\dag U_\mu^\dag \Gamma_\mu^R -2 \right),
\end{eqnarray}
the zero-th order term can be easily evaluated as
\begin{eqnarray}
H_0 &=& -\gamma_{d+1}\left[ a b(\hat p) 
+i\sum_\mu \gamma_\mu \sin(\hat p_\mu a)
\right] \equiv H_0(\hat p) , \\
(H_0)^2 &=&\left( a\omega (\hat p) \right)^2
\end{eqnarray}
where
\begin{eqnarray}
ab(\hat p) &\equiv& r\sum_\mu 
\left( 1-\cos(\hat p_\mu a)\right) -m_0 , \\
a\omega (\hat p) &\equiv& 
\sqrt{\sum_\mu \sin^2(\hat p_\mu a)
+\left( ab(\hat p)\right)^2}
\end{eqnarray}
and they satisfy
$a^2(\omega_{\hat p}^2 - b_{\hat p}^2)=\sum_\mu \sin^2(\hat p_\mu a)$.
\par
We expand the gauge field $A_\mu$ as 
\begin{equation}
A_\mu = \sum_k A_\mu (k) \hat\phi_k,
\end{equation}
where $A_\mu^*(-k)=A_\mu (k)$ since
$A_\mu^\dag = A_\mu$.
Then the first order term of $H$ becomes
\begin{equation}
H_1 = \sum_k \sum_\mu A_\mu (k) \hat\phi_k 
e^{-\frac{i}{2}k_\mu a}
\frac{\partial H_0 }
{\partial \hat p_\mu}\left(\hat p +\frac{k}{2} \right)
\end{equation}
where
\begin{equation}
\frac{\partial H_0}{\partial \hat p_\mu}(\hat p) = 
-ia \gamma_{d+1} 
\left[ \gamma_\mu \cos(\hat p_\mu a) 
-ri\sin(\hat p_\mu a) \right].
\end{equation}
No summation over $\mu$ is taken here.
\par
We now expand $(H^2)^{-\frac{1}{2}}$ 
for small gauge field configurations.
$(H^2)^{-\frac{1}{2}}$ can be expanded as 
\begin{eqnarray} 
&& \frac{1}{\sqrt{H^2}} = \int_t \frac{1}{t^2+H^2} 
= \int_t P_0 \sum_{m=0}^{\infty} (-1)^m
\left( \sum_{i=1}^{\infty} (H^2)_i \ P_0 \right)^m \n \\
&=& \int_t \left[ P_0 -P_0 (H^2)_1 P_0 -P_0 (H^2)_2 P_0
+P_0 \left((H^2)_1 P_0 \right)^2  \right.\n \\
&&\left. -P_0 (H^2)_3 P_0 
+P_0 \left\{(H^2)_1 P_0 ,(H^2)_2 P_0 \right\} 
-P_0 \left((H^2)_1 P_0 \right)^3 + \cdots \right]
\end{eqnarray}
where we introduced the abbreviated expressions such as 
$\int_t \equiv \frac{1}{\pi}\int_{-\infty}^{\infty} dt$ 
and $P_0 \equiv \frac{1}{t^2+(H^2)_0 }$.
Then we obtain the following expression for $\hat{\gamma}$
\begin{equation}
\hat{\gamma}=\frac{H}{\sqrt{H^2}} 
= \sum_{m=0}^\infty (-1)^m \int_t 
\left(\sum_{i=0}^{\infty}H_i \right) P_0  
\left( \sum_{j=1}^{\infty} (H^2)_j P_0 \right)^m .
\end{equation}
We are now interested in terms which contain 
$d/2$ gauge fields.  This gives the leading 
contribution to the chiral anomaly in $d$ dimensions. 
Since $(H)_i$ and $(H^2)_i$ contains at most 
one and two $\gamma$ matrices respectively, 
only the following term survives after taking
a spinor trace:
\begin{eqnarray}
&&(-1)^{d/2}\int_t H_0 P_0 
\left( (H^2)_1 P_0 \right)^{d/2} \n \\
&&=(-1)^{d/2}\int_t H_0 P_0 
\left(\{H_0,H_1 \}P_0 \right)^{d/2} \n \\
&&=(-1)^{d/2}\int_t H_0 P_0 \left( H_1 H_0 P_0 \right)^{d/2}.
\end{eqnarray}
The last equality follows from 
the fact that
the combination $H_0 P_0 H_0=H_0^2 P_0=P_0 H_0^2$ does 
not contain any $\gamma$ matrices and such a term vanishes after 
taking a spinor trace. 
Due to the same reason, all the other terms containing 
less or equal to the number of gauge fields $d/2$ vanish 
after taking a trace over spinors. 
Here in $d=2n$ dimensions, the other surviving terms contain
larger number of gauge fields.
\par
Therefore, if we take the leading order
term with $n$ gauge fields, 
the topological charge density becomes
\begin{eqnarray}
&&q(\lambda)
=\frac{1}{2}{\bold {Tr}}\left(\lambda \frac{H}{\sqrt{H^2}}\right) 
=\frac{1}{2N}\sum_p \Tr \hat\phi_p^\dag \lambda \frac{H}{\sqrt{H^2}} 
\hat\phi_p \n \\
&=&\frac{1}{2N}\sum_p \Tr \hat\phi_p^\dag \lambda 
\left[
\int_t  (-1)^n H_0 P_0 
\left( H_1 H_0 P_0 \right)^n 
\right]\hat\phi_p  
+{\cal O}\left(A_i^{\frac{d}{2}+1}\right) \n \\
&=&\frac{(-1)^n}{2N}\sum_p \sum_{k_0,k_1,\cdots,k_n}
\sum_{\mu_1,\cdots,\mu_n} \int_t 
\tr\left(\hat\phi_p^\dag \hat\phi_{k_0} \hat\phi_{k_n}\cdots
\hat\phi_{k_1} \hat\phi_p \right) \lambda(k_0) \n \\
&&\times A_{\mu_n}(k_n)\cdots A_{\mu_1}(k_1) 
e^{-\frac{i}{2}\sum_{i=1}^n (k_i)_{\mu_{i}}a} \n \\
&&
\times \tr_s\left[
\frac{H_0}{t^2+H_0^2}\left(p+\sum_{j_n=1}^n k_{j_n} \right) 
\cdot 
\frac{\partial H_0}{\partial p_{\mu_n}}
\left(p+\sum_{j_{n-1}=1}^{n-1} k_{j_{n-1}}+\frac{k_n}{2}\right) 
\cdot
\frac{H_0}{t^2+H_0^2}
\left(p+\sum_{h_{n-1}=1}^{n-1} k_{h_{n-1}} \right) \right. \n \\
&& 
\times \frac{\partial H_0}{\partial p_{\mu_{n-1}}}
\left(p+\sum_{j_{n-2}=1}^{n-2}k_{j_{n-2}} 
+\frac{k_{n-1}}{2}\right)
\cdots
\frac{H_0}{t^2+H_0^2}(p+k_1+k_2) \cdot
\frac{\partial H_0}{\partial p_{\mu_2}}
\left(p+k_1+\frac{k_2}{2}\right) \n \\
&& \times \left.
\frac{H_0}{t^2+H_0^2}(p+k_1) \cdot
\frac{\partial H_0}{\partial p_{\mu_1}}
\left(p+\frac{k_1}{2}\right) \cdot 
\frac{H_0}{t^2+H_0^2}(p)
\right] +{\cal O}\left(A_i^{\frac{d}{2}+1}\right) ,\label{finite}  
\end{eqnarray}
where the phase factor is given as
\begin{eqnarray}
&&\tr \hat\phi_{k_0} \hat\phi_{k_n}\cdots \hat\phi_{k_1} =N e^{i f(k,\theta)} 
\delta_{k_0, -\sum_{\lambda=1}^n k_\lambda} , \\
&& f(k,\theta) = \frac{1}{2} \sum_{i=0}^{n-1} 
\left( k_i^\alpha \sum_{j=i+1}^n k_j^\beta 
\theta_{\alpha \beta} \right). 
\end{eqnarray}

Now we take a large $L$ (continuum) limit
where $L_{phys}=La$ is fixed.
A requirement for the continuum limit
is that all the external momenta
$k_i$ are much smaller than the scale ${\cal O}(L)$. 
Since the noncommutative phase associated with
the star product is proportional to 
\begin{equation}
k_i^{\mu} k_{j}^{\nu} \theta_{\mu \nu}
=\frac{a L_{phys}}{2\pi} k_i^{\mu} k_{j}^{\nu} 
\epsilon_{\mu \nu},
\end{equation}
it survives if we scale the external momenta
as $k_i \propto L^{1/2}$ simultaneously.
Under this assumption, we can take a continuum 
limit while keeping the noncommutativity.
If we rescale the length so that the external momenta 
are of order $O(L^0)$, $L_{phys}$ scales as $O(L^{1/2})$. 
In this picture, the external momenta are
fixed but the size of the torus becomes very large and the 
geometry becomes noncommutative plane.

\par

In the continuum limit, we expand eq.(\ref{finite}) in terms of 
the external momenta $k_i^\mu$. 
Since each $p_\mu$ derivatives of $H_0$ is of order 
${\cal O}(a)$, we can take up to $n$ more $p_\mu$ derivatives 
so that eq.(\ref{finite}) survives in the continuum limit. 
After expanding the topological density 
in terms of the external momenta $k_i$
and taking a spinor trace, 
all the terms containing 
$H_0(p) (\partial H_0(p))^{m} H_0(p)$ 
vanish for an arbitrary odd integer $m$.
Hence, in the continuum limit, 
we need to take $n$ $p_\mu$ derivatives of all the $H_0$ in 
the numerators but the last one. 
Then $q(\lambda)$ becomes
\begin{eqnarray}
q(\lambda)&=&\frac{(-1)^n}{2N}\sum_p \sum_{k_0,k_1,\cdots,k_n}
\sum_{\mu_1,\cdots,\mu_n}  
\tr\left(\hat\phi_{k_0}\hat\phi_{k_n}\cdots
\hat\phi_{k_1} \right) \lambda(k_0) \n \\
&&\times A_{\mu_n}(k_n)\cdots A_{\mu_1}(k_1) 
\n \\
&&
\times \tr_s\left[ \left(\partial_{\nu_n} H_0 \right) 
\left(\partial_{\mu_n} H_0 \right)
\left(\partial_{\nu_{n-1}} H_0 \right)
\left(\partial_{\mu_{n-1}} H_0 \right) \cdots 
\left(\partial_{\nu_{1}} H_0 \right)
\left(\partial_{\mu_{1}} H_0 \right) H_0 \right] \n \\
&&\times \left(\sum_{j_n=1}^n k_{j_n}\right)_{\nu_n}
\left(\sum_{j_{n-1}=1}^{n-1} k_{j_{n-1}}\right)_{\nu_{n-1}} 
\cdots (k_1+k_2)_{\nu_2}(k_1)_{\nu_1}
\int_t \frac{1}{(t^2+H_0^2)^{n+1}} \n \\ 
&&+{\cal O}\left(A_i^{\frac{d}{2}+1}\right) . \label{kdel}
\end{eqnarray}
If we neglect the ordering of the plane waves, this
topological charge density becomes the same as 
the ordinary one in the lattice gauge theory. 
Hence it is clear now that the only difference
from the commutative case is that the
product of gauge fields is replaced with
the noncommutative star product.
\par
In order to evaluate the $p$ summation,
we follow the procedure in the paper \cite{Fujiwara:2002xh}.
First we introduce the following notations,
\begin{equation}
s_\mu=\sin{p_\mu a}, \quad c_\mu=\cos{p_\mu a} .
\end{equation}
By using the identities 
\begin{equation}
\tr \gamma_{d+1}\gamma_{\mu_1}\gamma_{\nu_1}\cdots
\gamma_{\mu_n}\gamma_{\nu_n}=i^n 2^n 
\epsilon_{\mu_1 \nu_1 \cdots \mu_n \nu_n}
\end{equation}
and
\begin{equation}
\int_t \frac{1}{(t^2 + c)^{n+1}}=
\frac{(2n-1)!!}{n! 2^n}\frac{1}{c^{n+\frac{1}{2}}} ,  
\end{equation}
and then taking a spinor trace, we have
\begin{eqnarray}
&&\tr_s\left[ (\partial_{\nu_n} H_0) 
(\partial_{\mu_n} H_0)
(\partial_{\nu_{n-1}} H_0)
(\partial_{\mu_{n-1}} H_0) \cdots 
(\partial_{\nu_{1}} H_0)
(\partial_{\mu_{1}} H_0) H_0
\right] \int_t \frac{1}{(t^2+H_0^2)^{n+1}} \n \\
&=&a^{2n}i^n \epsilon_{\nu_n \mu_n \cdots \nu_1 \mu_1}
I(p;m_0 ,r) \frac{(2n-1)!!}{n!},
\end{eqnarray}
where 
\begin{equation}
I(p;m_0,r)=\left(\prod_{\mu=1}^{d}c_\mu \right)
\left[m_0+r\sum_\rho (c_\rho -1) 
+r\sum_\rho \frac{s_\rho^2}{c_\rho} \right]
(H_0^2)^{-n-\frac{1}{2}} .
\end{equation}

In the continuum limit, the summation over the
momentum $p$ can be replaced by an integral as
\begin{equation}
\sum_p  = (L_{phys})^d \int \frac{d^d p}{(2 \pi)^d}.
\end{equation}
We thus have 
\begin{eqnarray} 
q(\lambda ) 
&=& \frac{1}{2}
\int \frac{d^d p}{(2\pi)^d}  (L_{phys})^d \sum_{k_1,\cdots,k_n}
\lambda \left(-\sum_{i=1}^n k_i \right) 
e^{i f(k,\theta)} \n \\
&&\times \epsilon_{\nu_n \mu_n \cdots \nu_1 \mu_1}
\left[(k_n)_{\nu_n}A_{\mu_n}(k_n) \right]
\cdots \left[(k_1)_{\nu_1}A_{\mu_1}(k_1)\right]
a^d I(p;m_0,r) 
\frac{(-i)^n (2n-1)!!}{n!} \n \\ 
&&+{\cal O}\left(A_i^{\frac{d}{2}+1}\right) . \label{Iphase}
\end{eqnarray}
We emphasize again that the above eq.(\ref{Iphase}) is the 
same as that in lattice gauge theory except for the existence 
of the phase factor 
$e^{if(k,\theta)}$.  
The evaluation of $I(p;m_0,r)$, which was done 
in ref.\cite{Fujiwara:2002xh}, is written in the appendix.
The result is summarized as
\begin{equation}
I(m_0,r)= a^d \int_{\cal B} d^d p \ I(p;m_0,r)=
\sum_{n_\pi=0}^{[m_0/2r]}(-1)^{n_\pi}{d\choose n_\pi}
\frac{2^{d+1}\pi^n n!}{d!}.
\end{equation}
We note here that $I(m_0,r)=\frac{2^{d+1}\pi^n n!}{d!}$ 
for $0<m_0/r<2$ and that the summation of $x$ can be replaced in 
the continuum limit as 
\begin{equation}
a^d \sum_x = \int d^d x .
\end{equation}
Then the final expression of the chiral anomaly is written as 
\begin{eqnarray} 
q(\lambda)
&=&\frac{1}{2} I(m_0,r) (L_{phys})^d 
\sum_{k_0,\cdots,k_n} \frac{1}{N^2} 
\sum_x e^{i(k_0 +\cdots + k_n)\cdot x} \lambda (k_0) 
e^{if(k,\theta)} \n \\
&&\times 
\epsilon_{\nu_n \mu_n \cdots \nu_1 \mu_1}
\left[(k_n)_{\nu_n}A_{\mu_n}(k_n) \right]
\cdots \left[(k_1)_{\nu_1}A_{\mu_1}(k_1)\right]
\frac{(-i)^n(2n-1)!!}{(2\pi)^d n!} 
+{\cal O}\left(A_i^{\frac{d}{2}+1}\right) \n \\
&=&\frac{(-1)^n (2n-1)!!}{2(2\pi)^d n!} I(m_0,r) 
\epsilon_{\nu_n \mu_n \cdots \nu_1 \mu_1}
\int d^d x \ \lambda (x) \star \partial_{\nu_n}A_{\mu_n}(x) 
\star \cdots \star \partial_{\nu_1}A_{\mu_1}(x) \n \\ 
&&+{\cal O}\left(A_i^{\frac{d}{2}+1}\right) .
\end{eqnarray}
In particular, when $0<m_0/r<2$ with which the overlap Dirac 
operator does not encounter the species doubling 
\cite{Neuberger:1998fp}, this reproduces the expected result.  

As we discussed in eqs.(\ref{gpsi})--(\ref{gnablapsi}), the action 
is invariant under gauge transformations and so is the 
topological charge if $\lambda$ transforms covariantly 
as in eq.(\ref{glambda}). Hence, the above $q(\lambda)$ must be 
also gauge invariant if higher order terms of the gauge 
fields are included. It will then become 
\begin{equation}
q(\lambda)=\frac{(-1)^n}{(4\pi)^n n!} 
\epsilon_{\nu_n \mu_n \cdots \nu_1 \mu_1} 
\int d^d x \ \lambda (x) \star F_{\nu_n \mu_n}(x) 
\star \cdots \star F_{\nu_1 \mu_1}(x)  
\end{equation}
for $0<m_0/r<2$ and 
$F_{\mu\nu}$ is a gauge covariant field strength 
in the continuum limit. 
This is the covariant form of the anomaly. 
In the case of gauge anomaly, similar calculation 
gives the same covariant form. A relation to the consistent 
form is discussed in refs.\cite{consistent}.

\section{Discussion}
In this paper we calculated 
the noncommutative chiral anomaly on arbitrary even dimensional
noncommutative  torus 
with overlap Dirac operator.
In the ``continuum limit'', we derived the correct Chern 
character including the star product.
\par
At the formal level, this looks very natural 
since the anomaly for the covariant chiral current 
in the fundamental representation
is diagramatically given by planar 
graphs\cite{Nishimura:2002hw,anomaly}.
But the topologies of the gauge configuration spaces
are very different between the ordinary lattice gauge theories
and its reduced models (or the noncommutative lattice theories), 
and further investigation is necessary to understand
the topological structure of 
gauge fields on noncommutative torus.
Namely, in the ordinary $U(1)$ gauge theories on the lattice
with $N^2$ lattice points, the gauge field configurations
have a topology of $\U(1)^{N^2 d}/\U(1)^{N^2}$
while  $\U(N)^d/\U(N)$ on the noncommutative torus.
In the lattice gauge theory the admissibility condition 
for the gauge field $\Vert 1- U(p) \Vert < \epsilon$ 
is imposed~\cite{Hernandez:1998et} 
so that the chirality 
operator and the GW Dirac operator are well-defined. 
The topological structure of gauge fields under the 
admissibility condition is 
classified explicitly, 
which enables us to obtain an anomaly free abelian 
chiral gauge theory on the lattice~\cite{Luscher:1998du}. 
It is interesting to investigate 
a similar condition on the noncommutative torus, 
with which the topological structure
for the gauge fields is defined. 
\par
Another related issue is the reduction of degrees of 
freedom in matrix models. In papers \cite{reduction},
the ordinary lattice gauge theories is embedded 
in matrix models by restricting the matrix 
degrees of freedom, and chiral anomaly and
the topological structure are discussed.
Their reduction is maximal. Namely,
they reduced $N^2$ degrees of freedom to $N$.
It is interesting and may be necessary
to consider weaker conditions of reductions 
for constructing well-defined topological 
structures in matrix models. 

\section*{Acknowledgements}
We would like to thank 
H. Aoki and J. Nishimura for discussions.

\appendix
%

\section{Evaluation of $I(m_0,r)$}
We now evaluate the following integral 
utilizing the degree of mapping 
following ref.\cite{Fujiwara:2002xh},
\begin{equation}
I(m_0,r)= a^d \int_{\cal B} d^d p I(p;m_0,r), 
\end{equation}
with
\begin{eqnarray}
I(p;m_0,r)&=&\left(\prod_{\mu=1}^{d}c_\mu \right)
\left\{\sum_\nu s_\nu^2 
+\left[m_0+r\sum_\nu (c_\nu -1)\right]^2 
\right\}^{-n-\frac{1}{2}} \n \\
&&\times \left[m_0+r\sum_\rho (c_\rho -1) 
+r\sum_\rho \frac{s_\rho^2}{c_\rho} \right] 
%
\end{eqnarray}
and
\begin{equation}
{\cal B}=\left\{k_\mu\in{\mathbb R}^d\bigm|
-{\pi\over2a}\leq k_\mu\leq{3\pi\over2a}\right\}.
\label{twentythree}
\end{equation}
Here we have shifted the integration region
as~$k_\mu\in[-\frac{\pi}{a},\frac{\pi}{a}]\to k_\mu
\in[-\frac{\pi}{2a},\frac{3\pi}{2a}]$ for later convenience.


We introduce a mapping from the
Brillouin zone~${\cal B}$ to the unit sphere~$S^d$. 
The mapping is defined by
\begin{eqnarray}
&&\theta_0=m_0+r\sum_\mu(c_\mu -1),
\label{twentyfour}\\
&&\theta_\mu=s_\mu,\qquad{\rm for}\qquad\mu=1,\ldots,d,
\label{twentyfive}
\end{eqnarray}
and
\begin{equation}
x_I={\theta_A\over\epsilon},\qquad\epsilon=
\sqrt{\sum_I\theta_I^2},
\label{twentysix}
\end{equation}
where $x_I$ ($I=0$, $1$, $\ldots$, $d$) is the coordinate
of~${\mathbb R}^{d+1}$ in which the unit 
sphere~$\sum_Ix_I^2=1$ is embedded.
$p_\mu\to x_I$ defines a mapping~$f:T^d\to S^d$. 
The crucial observation is that the volume form on this
sphere coincides with the integrand of~$I(m_0,r)$:
\begin{eqnarray}
\Omega&=&{1\over d!}\,
\epsilon_{A_0\cdots A_d}x_{A_0}\rmd x_{A_1}\wedge\cdots\wedge\rmd x_{A_d}
\nonumber\\
&=&{1\over d!}{1\over\epsilon^{d+1}}\,\epsilon_{A_0\cdots A_d}
\theta_{A_0}\rmd\theta_{A_1}\wedge\cdots\wedge\rmd\theta_{A_d}
\nonumber\\
&=&a^d I(p;m_0,r)\,\rmd p_1\wedge\cdots\wedge\rmd p_d.
\label{twentyseven}
\end{eqnarray}
This shows that the integral of~$\Omega$ on a (sufficiently small) coordinate
patch~$U$ on~$S^d$ is given by
\begin{equation}
\int_U\Omega=\sgn\left[I(p^j;m_0,r)\right]
\int_{U^j} a^d I(p;m_0,r)\,\rmd 
p_1\wedge\cdots\wedge\rmd p_d,
\label{twentyeight}
\end{equation}
where $U^j$ ($j=1$, \dots, $m$) is a component of the inverse image of~$U$
under~$f$, $f^{-1}(U)$:
\begin{equation}
f^{-1}(U)=U^1\cup\cdots\cup U^m\subset T^d,
\label{twentynine}
\end{equation}
and $p^j$ ($j=1$, \dots, $m$) is a certain point on~$U^j$. For a sufficiently
small~$U$, $U_j$ are pairwise disjoint. We take preimages of a point~$y\in U$
under~$f$, $f^{-1}(y)$, as $p^j$. Then by summing both sides
of~eq.~(\ref{twentyeight}) over~$j$, we have\footnote{In deriving this
relation, we have assumed that $U$ is within the range of~$f$, i.e., the
inverse image~$f^{-1}(U)$ is not empty. This relation itself, however, is
meaningful even if $U$ is not within the range of~$f$, if one sets $\deg f=0$
for such case. As a consequence, eq.~(\ref{thirtytwo}) holds even if
$f:T^d\to S^d$ is not a surjection, i.e., not an onto-mapping.}
\begin{equation}
\sum_j \int_{U^j} a^d I(p;m_0,r)\,\rmd p_1\wedge\cdots\wedge
\rmd p_d=(\deg f)\int_U\Omega,
\label{thirty}
\end{equation}
where the degree of the mapping~$f$ is given by
\begin{equation}
\deg f=\sum_{f(p^j)=y}\sgn\left[I(p^j;m_0,r)\right].
\label{thirtyone}
\end{equation}
In general, the degree of the mapping $f:T^d\to S^d$ is defined by a sum of the
signature of jacobian of the coordinate transformation between~$T^d$
and~$S^d$ over preimages of a point $y\in S^d$. An important mathematical
fact is that the degree takes the same value for all
coordinate patches~$U$ of~$S^d$. We thus have
\begin{equation}
I(m_0,r)
=\int_{T^d} a^d I(p;m_0,r)\,\rmd p_1\wedge\cdots\wedge\rmd p_d
=(\deg f)\int_{S^d}\Omega,
\label{thirtytwo}
\end{equation}
where $\int_{S^d}\Omega$ is given by the volume of the unit sphere~$S^d$:
\begin{equation}
\int_{S^d}\Omega=\vol(S^d)={2^{d+1}\pi^nn!\over d!}.
\label{thirtythree}
\end{equation}

We may choose any point $y$ on~$S^d$ to evaluate the degree~(\ref{thirtyone}).
We choose $y=(1,0,\cdots,0)$. This requires
\begin{equation}
x_\mu(p^j)=
{s_\mu\over\epsilon}=0,\qquad{\rm for}\qquad\mu=1,\ldots,d,
\label{thirtyfour}
\end{equation}
and
\begin{equation}
x_0(p^j)=
{m_0+r\sum_\mu(c_\mu-1)\over\left|m_0+r\sum_\mu(c_\mu-1)\right|}
=1.
\label{thirtyfive}
\end{equation}
Note that eq.~(\ref{thirtyfive}) is equivalent to the
condition~$m_0/r+\sum_\mu(c_\mu-1)>0$. Now eq.~(\ref{thirtyfour}) implies that
$p_\mu^j=0$ or~$\pi$ for each direction~$\mu$. We denote the number of $\pi$'s
appearing in $p^j$ by an integer~$n_\pi\geq0$,
\begin{equation}
p^j=(\underbrace{\pi,\ldots,\pi}_{n_\pi},0,\ldots,0),
\label{thirtysix}
\end{equation}
irrespective of the position of $\pi$'s. For a given~$n_\pi$, the number of
such $p^j$ is ${d\choose n_\pi}$. The second relation~(\ref{thirtyfive}) on the
other hand requires~$n_\pi<m_0/(2r)$. At those $p^j$, we have
\begin{equation}
\sgn\left[I(p^j;m_0,r)\right]=\prod_\mu c_\mu=(-1)^{n_\pi}.
\label{thirtyseven}
\end{equation}
Thus eq.~(\ref{thirtyone}) gives
\begin{equation}
\deg f=\sum_{n_\pi=0}^{[m_0/2r]}{d\choose n_\pi}(-1)^{n_\pi}.
\label{thirtyeight}
\end{equation}

Combining eqs.~(\ref{thirtytwo}), (\ref{thirtythree})
and~(\ref{thirtyeight}), we finally obtain
\begin{equation}
I(m_0,r)=\sum_{n_\pi=0}^{[m_0/2r]}(-1)^{n_\pi}{d\choose n_\pi}
\frac{2^{d+1}\pi^n n!}{d!}.\label{finalI}
\end{equation}


\begin{thebibliography}{99}


\bibitem{IKKT}
N.~Ishibashi, H.~Kawai, Y.~Kitazawa and A.~Tsuchiya,
\NPB{498,1997,467} 
[hep-th/{9612115}];
H.~Aoki, S.~Iso, H.~Kawai, Y.~Kitazawa, A.~Tsuchiya and T.~Tada,
\PTPS{134,1999,47} 
[hep-th/{9908038}].

\bibitem{spacetime}
H.~Aoki, S.~Iso, H.~Kawai, Y.~Kitazawa and T.~Tada,
\PTP{99,1998,713}
[hep-th/{9802085}];
J.~Nishimura and F.~Sugino,
\JHEP{0205,2002,001}
[hep-th/{0111102}];
H.~Kawai, S.~Kawamoto, T.~Kuroki, T.~Matsuo and S.~Shinohara,
\NPB{647,2002,153}
[hep-th/{0204240}].
%

\bibitem{Connes}
A. Connes, Noncommutative geometry, Academic Press, 1990.


\bibitem{CDS}
A.~Connes, M.~R.~Douglas and A.~Schwarz,
\JHEP{9802,1998,003}
[hep-th/{9711162}].


\bibitem{Li}
M.~Li,
\NPB{499,1997,149}
[hep-th/{9612222}].

\bibitem{AIIKKT}
H.~Aoki, N.~Ishibashi, S.~Iso, H.~Kawai, Y.~Kitazawa and T.~Tada,
\NPB{565,2000,176}
[hep-th/{9908141}].

\bibitem{Ambjorn}
J.~Ambjorn, Y.~M.~Makeenko, J.~Nishimura and R.~J.~Szabo,
\JHEP{9911,1999,029}
[hep-th/{9911041}]; 
\PLB{480,2000,399}
[hep-th/{0002158}]; 
\JHEP{0005,2000,023}
[hep-th/{0004147}].


\bibitem{NCYM}
N.~Ishibashi, S.~Iso, H.~Kawai and Y.~Kitazawa,
\NPB{573,2000,573}
[hep-th/{9910004}];  
S.~Minwalla, M.~Van Raamsdonk and N.~Seiberg,
\JHEP{0002,2000,020}
[hep-th/{9912072}];
S.~Iso, H.~Kawai and Y.~Kitazawa,
\NPB{576,2000,375}
[hep-th/{0001027}]; 
A.~Matusis, L.~Susskind and N.~Toumbas,
\JHEP{0012,2000,002}
[hep-th/{0002075}];
N.~Ishibashi, S.~Iso, H.~Kawai and Y.~Kitazawa,
\NPB{583,2000,159}
[hep-th/{0004038}].

\bibitem{metric}
S.~Iso and H.~Kawai,
Int.\ J.\ Mod.\ Phys.\ A {\bf 15} (2000), 651
[hep-th/{9903217}];
T.~Azuma, S.~Iso, H.~Kawai and Y.~Ohwashi,
\NPB{610,2001,251}
[hep-th/{0102168}];
S.~Iso, Y. Kimura, K. Tanaka and K. Wakatsuki,
\NPB{604,2001,121}
[hep-th/{0101102}].

\bibitem{AIS}
H.~Aoki, S.~Iso and T.~Suyama,
\NPB{634,2002,71}
[hep-th/{0203277}].


\bibitem{AIN2}
H.~Aoki, S.~Iso and K.~Nagao,
hep-th/{0209223}.

\bibitem{GW}P.~H.~Ginsparg and K.~G.~Wilson,
\PRD{25,1982,2649}.

\bibitem{NV}
J.~Nishimura and M.~A.~Vazquez-Mozo,
\JHEP{0108,2001,033}
[hep-th/{0107110}].


\bibitem{dwfermion}
D.~B.~Kaplan,
\PLB{288,1992,342}
[hep-lat/{9206013}];

Y.~Shamir,
\NPB{406,1993,90}
[hep-lat/{9303005}].

\bibitem{Nagao:2001tc}
K.~Nagao,
\NPB{636,2002,264}
[hep-lat/{0112030}].

\bibitem{Neuberger:1998fp}H.~Neuberger,
\PLB{417,1998,141}
[hep-lat/{9707022}];
\PRD{57,1998,5417}
[hep-lat/{9710089}];
\PLB{427,1998,353}
[hep-lat/{9801031}].


\bibitem{Luscher:1998du}
M.~L\"uscher,
\NPB{549,1999,295} [hep-lat/{9811032}].


\bibitem{Kikukawa:1998pd}
Y.~Kikukawa and A.~Yamada,
\PLB{448,1999,265} [hep-lat/{9806013}]; 
%
K.~Fujikawa,
\NPB{546,1999,480} [hep-th/{9811235}]; 
%
D.H.~Adams,
\ANN{296,2002,131} [hep-lat/{9812003}];
\JMP{42,2001,5522} [hep-lat/{0009026}]; 
%
H.~Suzuki,
\PTP{102,1999,141} [hep-th/{9812019}]; 
%
T.-W.~Chiu and T.-H.~Hsieh,
hep-lat/{9901011};
\PRD{65,2002,054508}
[hep-lat/{0109016}]; 
%
T.~Reisz and H.J.~Rothe,
\PLB{455,1999,246} [hep-lat/{9903003}]; 
%
M.~Frewer and H.J.~Rothe,
\PRD{63,2001,054506}
[hep-lat/{0004005}]; 
%
K.~Fujikawa and M.~Ishibashi,
\NPB{587,2000,419} [hep-lat/{0005003}]. 


\bibitem{Fujiwara:2002xh}
T.~Fujiwara, K.~Nagao and H.~Suzuki,
\JHEP{0209,2002,025}
[hep-lat/{0208057}].


\bibitem{Luscher:1999kn}
M.~L\"uscher,
\NPB{538,1999,515} [hep-lat/{9808021}]; 
%
T.~Fujiwara, H.~Suzuki and K.~Wu,
\NPB{569,2000,643} [hep-lat/{9906015}];
\PLB{463,1999,63} [hep-lat/{9906016}]; 
%
H.~Igarashi, K.~Okuyama and H.~Suzuki,
hep-lat/{0206003}.


%
\bibitem{Correa:2001rd}
D.~H.~Correa and E.~F.~Moreno,
\PLB{534,2002,185}
[arXiv:hep-th/0111054].


\bibitem{Hernandez:1998et}
P.~Hernandez, K.~Jansen and M.~L\"uscher,
\NPB{552,1999,363}
[hep-lat/{9808010}].

%
\bibitem{Nishimura:2002hw}
J.~Nishimura and M.~A.~Vazquez-Mozo,
arXiv:hep-lat/0210017.


\bibitem{Luscher}M.~L\"uscher,
\PLB{428,1998,342} 
[hep-lat/{9802011}].
 %
\bibitem{Nieder}F.~Niedermayer,
\NPB{73,1999,105}
[hep-lat/{9810026}].

%
\bibitem{anomaly}
F.~Ardalan and N.~Sadooghi,
Int.\ J.\ Mod.\ Phys.\ A {\bf 16} (2001), 3151
[hep-th/{0002143}];
J.~M.~Gracia-Bondia and C.~P.~Martin,
\PLB{479,2000,321}
[hep-th/{0002171}];
F.~Ardalan and N.~Sadooghi,
Int.\ J.\ Mod.\ Phys.\ A {\bf 17} (2002), 123
[hep-th/{0009233}];
%
R.~Banerjee and S.~Ghosh,
Phys.\ Lett.\ B {\bf 533}, 162 (2002)
[arXiv:hep-th/0110177];
%
A.~Armoni, E.~Lopez and S.~Theisen,
\JHEP{0206,2002,050}
[hep-th/{0203165}]; 
H.~Aoki, S.~Iso and K.~Nagao,
hep-th/{0209137}.
%
\bibitem{Chandrasekharan:1998wg}
S.~Chandrasekharan,
\PRD{60,1999,074503}
[hep-lat/{9805015}]; 
%
I.~Ichinose and K.~Nagao,
Chin.\ J.\ Phys.\ {\bf 38} (2000), 671  
[hep-lat/{9912011}]; 
%
Mod.\ Phys.\ Lett.\ A {\bf 15} (2000), 857 ; 
%
\NPB{577,2000,279}
[hep-lat/{9910031}]; 
%
\NPB{596,2001,231}
[hep-lat/{0008002}]; 
%
M.~Golterman and Y.~Shamir,
\JHEP{0009,2000,006}
[hep-lat/{0007021}]; 
%
L.~Giusti, G.~C.~Rossi, M.~Testa and G.~Veneziano,
\NPB{628,2002,234}
[hep-lat/{0108009}].


%
\bibitem{reduction} 
J.~Kiskis, R.~Narayanan and H.~Neuberger,
\PRD{66,2002,025019}
[hep-lat/{0203005}];
Y.~Kikukawa and H.~Suzuki,
\JHEP{0209,2002,032}
[hep-lat/{0207009}].
%

\bibitem{consistent}
L.~Bonora, M.~Schnabl and A.~Tomasiello,
\PLB{485,2000,311}
[hep-th/{0002210}];
%
E.~F.~Moreno and F.~A.~Schaposnik,
\JHEP{0003,2000,032}
[hep-th/{0002236}]; 
%
\NPB{596,2001,439}
[hep-th/{0008118}]; 
%
C.~P.~Martin,
J.\ of Phys.\ A {\bf 34} (2001), 9037
[hep-th/{0008126}]; 
%
M.~T.~Grisaru and S.~Penati,
\PLB{504,2001,89}
[hep-th/{0010177}];
%
L.~Bonora and A.~Sorin,
\PLB{521,2001,421}
[hep-th/{0109204}]; 
C.~P.~Martin,
\NPB{623,2002,150}
[hep-th/{0110046}].


\end{thebibliography}
\end{document}